\documentclass[twocolumn]{aastex63}

\usepackage{ulem}

\submitjournal{ApJ}

\shorttitle{Manganese Line in Cassiopeia A}
\shortauthors{Sato et al.}

\begin{document}

\title{A Sub-Solar Metallicity Progenitor for Cassiopeia A, the remnant of a Type IIb Supernova} 

\correspondingauthor{Toshiki Sato}
\email{toshiki.sato@riken.jp}

\author[0000-0001-9267-1693]{Toshiki Sato}
\affil{RIKEN, 2-1 Hirosawa, Wako, Saitama 351-0198, Japan}
\affil{NASA, Goddard Space Flight Center, 8800 Greenbelt Road, Greenbelt, MD 20771, USA}
\affil{Department of Physics, University of Maryland Baltimore County, 1000 Hilltop Circle, Baltimore, MD 21250, USA}

\author[0000-0002-8967-7063]{Takashi Yoshida}
\affil{Department of Astronomy, Graduate School of Science, University of Tokyo, 7-3-1 Hongo, Bunkyo-ku, Tokyo 113-0033, Japan}

\author{Hideyuki Umeda}
\affil{Department of Astronomy, Graduate School of Science, University of Tokyo, 7-3-1 Hongo, Bunkyo-ku, Tokyo 113-0033, Japan}

\author[0000-0002-7025-284X]{Shigehiro Nagataki}
\affiliation{Astrophysical Big Bang Laboratory (ABBL), RIKEN Cluster for Pioneering Research, 2-1 Hirosawa, Wako, Saitama 351-0198, Japan}
\affiliation{RIKEN Interdisciplinary Theoretical and Mathematical Science Program (iTHEMS), 2-1 Hirosawa, Wako, Saitama 351-0198, Japan}

\author[0000-0002-7025-284X]{Masaomi Ono}
\affiliation{Astrophysical Big Bang Laboratory (ABBL), RIKEN Cluster for Pioneering Research, 2-1 Hirosawa, Wako, Saitama 351-0198, Japan}
\affiliation{RIKEN Interdisciplinary Theoretical and Mathematical Science Program (iTHEMS), 2-1 Hirosawa, Wako, Saitama 351-0198, Japan}

\author[0000-0003-2611-7269]{Keiichi Maeda}
\affiliation{Department of Astronomy, Kyoto University, Kitashirakawa-Oiwake-cho, Sakyo-ku, Kyoto 606-8502, Japan}

\author[0000-0002-8032-8174]{Ryosuke Hirai}
\affiliation{OzGrav: The Australian Research Council Centre of Excellence for Gravitational Wave Discovery, Clayton, VIC 3800, Australia}
\affiliation{School of Physics and Astronomy, Monash University, VIC 3800, Australia}

\author[0000-0002-8816-6800]{John P. Hughes}
\affiliation{Department of Physics and Astronomy, 
Rutgers University, 136 Frelinghuysen Road, Piscataway, 
NJ 08854-8019, USA}

\author[0000-0003-2063-381X]{Brian J. Williams}
\affiliation{NASA, Goddard Space Flight Center, 8800 Greenbelt Road, Greenbelt, MD 20771, USA}

\author{Yoshitomo Maeda}
\affiliation{Institute of Space and Astronautical Science, Japan Aerospace Exploration Agency 3-1-1 Yoshinodai, Chuo-ku, Sagamihara, Kanagawa 229-8510}
\affiliation{The Graduate University for Advanced Studies, 3-1-1 Yoshinodai, Chuo-ku, Sagamihara, Kanagawa 252-5210, Japan}


\begin{abstract}

We report, for the first time, the detection of the Mn-K$\alpha$ line in the Type IIb supernova (SN IIb) remnant, Cassiopeia A. Manganese ($^{55}$Mn after decay of $^{55}$Co), a neutron-rich element, together with chromium ($^{52}$Cr after decay of $^{52}$Fe), is mainly synthesized in core-collapse supernovae at the explosive incomplete Si burning regime. Therefore, the Mn/Cr mass ratio with its neutron excess reflects the neutronization at the relevant burning layer during the explosion.  {\it Chandra}'s deep archival X-ray data of Cassiopeia A indicate a low Mn/Cr mass ratio with values in the range 0.10--0.66, which, when compared to one-dimensional SN explosion models, requires that the electron fraction be 0.4990 $\lesssim Y_{\rm e} \lesssim$ 0.5 at the incomplete Si burning layer. An explosion model assuming a solar-metallicity progenitor with a typical explosion energy ($1 \times 10^{51}$ erg) fails to reproduce such a high electron fraction. In such models, the explosive Si-burning regime extends only to the Si/O layer established during the progenitor's hydrostatic evolution; the $Y_e$ in the Si/O layer is lower than the value required by our observational constraints.  We can satisfy the observed Mn/Cr mass ratio if the explosive Si-burning regime were to extend into the O/Ne hydrostatic layer, which has a higher $Y_{\rm e}$. This would require an energetic ($> 2 \times 10^{51}$ erg) and/or asymmetric explosion of a sub-solar metallicity progenitor ($Z \lesssim 0.5Z_{\odot}$) for Cassiopeia A. The low initial metallicity can be used to rule out a single-star progenitor, leaving the possibility of a binary progenitor with a compact companion (white dwarf, neutron star or black hole). We discuss the detectability of X-rays from Bondi accretion onto such a compact companion around the explosion site.  We also discuss other possible mass-loss scenarios for the progenitor system of Cassiopeia A.


\end{abstract}


\section{Introduction} \label{sec:intro}

Cassiopeia A is one of the most well-studied Galactic supernova remnants (SNRs). Because of its young age ($\sim$350 yrs) and close distance (3.4 kpc) \citep{2006ApJ...645..283F}, the remnant provides a unique opportunity to investigate the death of a massive star and to test theoretical models of core-collapse supernovae (CC SNe) \citep[e.g.,][]{1998ApJ...492L..45N,2016ApJ...822...22O,2017ApJ...842...13W}. 

\begin{figure*}[t!]
 \begin{center}
  \includegraphics[width=16cm, bb=0 0 904 421]{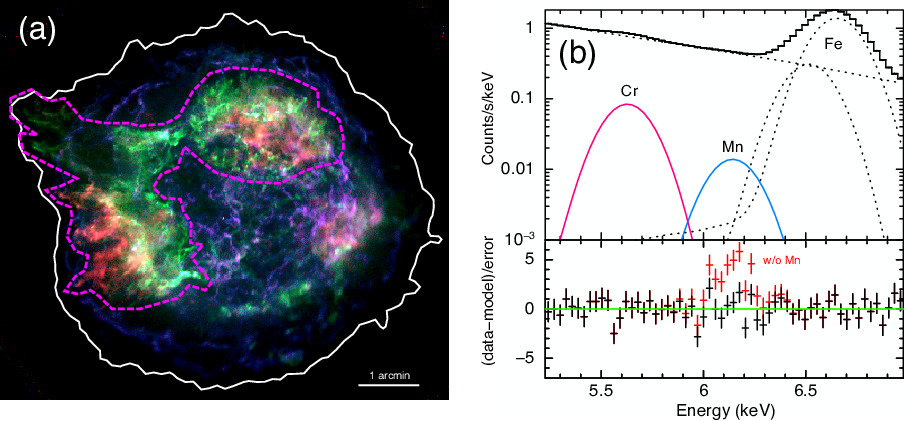}
 \end{center}
\caption{(a) {\it Chandra} X-ray image of Cassiopeia A, combining images in green, red and blue made from energy bands of 1.76--1.94 keV (Si K$\alpha$), 6.54--6.92 keV (Fe K$\alpha$) and 4.2--6.0 keV (continuum), respectively. White and magenta contours show the entire remnant and thermal dominant regions, respectively. (b) The {\it Chandra} X-ray spectrum in 5.2--7.0 keV band in the thermal dominant region. The broken and solid lines show the best-fit model consisting of power law + several Gaussian models. The magenta and sky blue lines show the Cr K$\alpha$ and Mn K$\alpha$ lines, respectively. In this study, all the widths of the Gaussian models are linked to each other.}
\label{fig:f1}
\end{figure*}

The current consensus on the origin of Cassiopeia A is a Type IIb SN (SN IIb) where the progenitor star's hydrogen envelope was mostly stripped by a binary interaction. The SN IIb classification is strongly supported by its light echo spectrum \citep{2008Sci...320.1195K}; it showed a close resemblance to the spectrum of the prototypical SN IIb 1993J. SN 1993J is thought to be the supernova of a red supergiant star with a main sequence mass of 13--20 $M_\odot$ \citep{1993Natur.364..507N,1994ApJ...429..300W}, whose hydrogen envelope had been lost to a slightly less massive close binary companion star. 
In addition, \cite{2006ApJ...640..891Y} independently inferred a 15--25 $M_{\odot}$ progenitor based on the amount of $^{44}$Ti and $^{56}$Ni in Cassiopeia A. Thus, the similarity with SN 1993J and the estimated mass range of the progenitor imply some difficulty for the loss of its envelope only through the action of its own stellar wind \citep{2003ApJ...591..288H}. However, no  surviving companion star, that should exist in almost all cases of binary SN-IIb progenitors, has been found at the explosion site  \citep{2018MNRAS.473.1633K,2019A&A...623A..34K}, challenging the binary interaction scenario. 

To understand the mass-loss history of SN~IIb progenitors, the initial metallicity of progenitor stars, which is directly related to the mass-loss rates by stellar winds, is one of the most important parameters. On the other hand, at present, there is no solid method for estimating the initial metallicity of CC SNRs such as Cassiopeia A. In the case of Type Ia SNRs, the mass ratio between manganese and chromium, which are produced together at the incomplete Si burning regime and have different neutron excess, can be a good tracer of the initial metallicity \citep[][]{2008ApJ...680L..33B,2013ApJ...767L..10P,2020ApJ...890..104S}. Here, the elements processed by the CNO cycle during stellar evolution pile up into $^{14}$N, which is burnt to $^{22}$Ne in the He burning stage through the reactions $^{14}$N($\alpha$,$\gamma$)$^{18}$F($\beta^+$)$^{18}$O($\alpha$,$\gamma$)$^{22}$Ne. The $\beta^+$ decay in this process increases the neutron excess \citep{2003ApJ...590L..83T}. Such neutronization affected by the initial metallicity also occurs in the stellar evolution of massive stars \citep[e.g.,][]{1985ApJ...295..604T}. Thus, the Mn/Cr ratio should be sensitive to the initial metallicity of CC SNRs, too. However, the electron fraction (electron-to-baryon number ratio), an indicator of the neutron excess, at layers experiencing Si-burning in CC SNe is further complicated by other effects (e.g., electron capture during the stellar evolution, neutrino heating during the explosion). Therefore, to discuss the initial metallicity using the Mn/Cr ratio in CC SNRs, we need to calculate nuclear reactions comprehensively through the progenitor's hydrostatic evolution to explosion. 


For Cassiopeia A, the Mn-K line has not been detected previously in X-rays, although the Cr-K line ($\sim$5.6 keV) has \citep[e.g.,][]{2009PASJ...61.1217M,2009ApJ...692..894Y,2013ApJ...766...44Y}. In this paper, we report the detection of Mn-K line in Cassiopeia A for the first time. Using the Mn/Cr ratio, we discuss the degree of neutronization at the interior of the progenitor star of Cassiopeia A. Comparing with one-dimensional explosive nucleosynthesis models for CC SNe, we provide the first constraint on the initial metallicity, which will help us to understand the mass-loss history of the SN-IIb progenitor. 

Section 2 details the X-ray analysis that has resulted in the Mn line detection.  In the section following that we convert the measured Mn/Cr flux ratios to mass ratios and relate them to the expected neutron excess required by the material to be burnt during the explosion.  In section 4, the neutron excess value is used to explore what constraints can be set on the initial metallicity of the progenitor star and the energy of the explosion using 1D stellar evolution models for progenitors with main sequence masses in the range 13--20 $M_{\odot}$.  Section 5 ties the results of section 4 to other information about Cassiopeia A to help understand how the progenitor star lost its hydrogen envelope before exploding. One scenario involves a binary with a compact object that may have left an isolated runaway black hole within the remnant's interior; the detectability of such an object is the focus of section 6.  The final section summarizes the key points of our study.

\section{Detection of the Manganese line in Cassiopeia A} \label{sec:intro}
{\it Chandra} ACIS-S has observed Cassiopeia A several times since launch
\citep[e.g.,][]{2000ApJ...528L.109H,2000ApJ...537L.119H,2004ApJ...615L.117H,2011ApJ...729L..28P,2012ApJ...746..130H,2014ApJ...789..138P,2017ApJ...836..225S,2018ApJ...853...46S}.
We used all ACIS-S observations from 2000--2018 with a total exposure of $\sim$1.57 Msec.
We reprocessed the event files (from level 1 to level 2) to remove pixel randomization
and to correct for CCD charge transfer efficiencies using CIAO \citep{2006SPIE.6270E..1VF} version 4.6 and CalDB 4.6.3. The bad grades were filtered out and good time intervals were reserved. We also used the {\it Suzaku} observation of this remnant in 2012 with the exposure time of $\sim$102 ksec to compare with the ACIS-S observations.


Figure \ref{fig:f1}(a) shows the {\it Chandra} image of Cassiopeia A. We extract the X-ray spectra from the entire remnant (white contour) and the thermal dominant region (magenta contour). To detect a faint line like the Mn K$\alpha$, the continuum emission is an obstacle. Thus, we chose the thermal dominant region while avoiding the bright non-thermal X-rays (blue color in the figure).

We fitted the X-ray spectra with a spectral model consisting of a power law and four Gaussian models (Figure \ref{fig:f1}(b)). As a result, we found a statistically-significant line structure at $\sim$6.16 keV in Cassiopeia A (Table \ref{tab:Gafit}). In particular, the spectrum in the thermal dominant region showed a high significance level ($\Delta \chi^2 = 35.3$ for one degree of freedom, F-test probability\footnote{see  https://heasarc.gsfc.nasa.gov/xanadu/xspec/manual/node83.html} = 5.4$\times10^{-5}$) of the Mn detection thanks to the low level of the continuum emission (Table \ref{tab:Gafit}). In the {\it Suzaku} observation of the entire remnant, it was difficult to determine the centroid energy, thus we fixed it to 6.16 keV. Although the flux errors in the {\it Suzaku} data are larger because of the shorter exposure time, the Mn/Cr flux ratios in all the observations are consistent with each other, which enhances the robustness of our measurements.

\begin{table}[t]
\caption{Summary of Gaussian-model fits for the Mn and Cr lines in Cassiopeia A using {\it Chandra} and {\it Suzaku}.}
\begin{center}
\begin{tabular}{lccc}
\hline
Emission        & Centroid                      & Flux                              & significance\\ 
                & (keV)                         & (10$^{-4}$ ph/cm$^{2}$/s) \\ \hline
\multicolumn{4}{l}{{\bf \textit{Chandra}} (entire remnant)}\\ 
Cr K$\alpha$    &  5.632$^{+0.006}_{-0.005}$    & 1.40$^{+0.06}_{-0.07}$& 36 $\sigma$\\
Mn K$\alpha$    &  6.16$^{+0.04}_{-0.03}$       & 0.29$^{+0.06}_{-0.11}$& 3.9 $\sigma$\\ 
$F_{\rm Mn}$/$F_{\rm Cr}$ & ---                 & 0.16$^{+0.05}_{-0.08}$& ---\\
\multicolumn{4}{l}{{\bf \textit{Chandra}} (thermal dominant)}\\ 
Cr K$\alpha$    &  5.637$^{+0.006}_{-0.005}$    & 1.02$^{+0.07}_{-0.03}$& 41 $\sigma$ \\
Mn K$\alpha$    &  6.15$^{+0.02}_{-0.03}$       & 0.23$^{+0.08}_{-0.06}$& 6.4 $\sigma$\\ 
$F_{\rm Mn}$/$F_{\rm Cr}$ & ---                 & 0.22$^{+0.08}_{-0.06}$& ---     \\
\multicolumn{4}{l}{{\bf \textit{Suzaku}} (entire remnant)}\\ 
Cr K$\alpha$    &  5.62$^{+0.02}_{-0.01}$       & 1.6$^{+0.2}_{-0.3}$ & 10 $\sigma$\\
Mn K$\alpha$    &  6.16 (fix)                   & 0.3$^{+0.3}_{-0.2}$ & 2.6 $\sigma$\\ 
$F_{\rm Mn}$/$F_{\rm Cr}$ & ---                 & 0.19$^{+0.20}_{-0.12}$& ---     \\ \hline
\end{tabular}
\label{tab:Gafit}
\end{center}
\end{table}

The centroid energy of $\sim$6.16 keV corresponds to the Mn K$\alpha$ line for the ionizing plasma whose plasma parameters are consistent with those for Cassiopeia A \citep[i.e., in the box of Figure \ref{fig:f2};][]{2002A&A...381.1039W,2012ApJ...746..130H}. The Mn is synthesized as $^{55}$Co together with the Cr ($^{52}$Fe decays to the stable $^{52}$Cr) at the incomplete Si burning regime. Therefore, we assumed the plasma parameters are in common between them.


\section{The Mn/Cr ratio and Electron Fraction}

The amount of Mn is sensitive to the neutron excess at the incomplete Si burning layer. We note that the products of Si-burning do not depend sensitively on the different compositions of the layers. Here, the yield of Mn becomes an excellent tracer of the neutronization during the explosion. The degree of neutronization at the incomplete Si burning mainly depends on the initial metallicity and the electron capture in the progenitor star \citep[e.g.,][]{1985ApJ...295..604T,1996ApJ...460..408T}. Therefore, the observational constraint of the degree of neutronization would be useful to extract the progenitor information. \cite{2005ApJ...619..427U} investigated abundance of Fe-peak elements as a function of the electron fraction (electron-to-baryon number ratio), $Y_{\rm e}$, in the Si-burning region (see Fig. 4 in this paper). Around $Y_{\rm e} \approx 0.5$, the yields of Mn drastically changes while that of Cr shows a flat evolution.

From the Mn/Cr flux ratio in Table \ref{tab:Gafit}, we could estimate its mass ratio: $M_{\rm Mn}/M_{\rm Cr}$ $=$  1.058 $\times$ ($F_{\rm Mn}/F_{\rm Cr}$)/($\epsilon_{\rm Mn}/\epsilon_{\rm Cr}$), where 1.058 is the ratio of atomic masses, $F_{\rm Mn}/F_{\rm Cr}$ is the line flux ratio, and $\epsilon_{\rm Mn}/\epsilon_{\rm Cr}$ is the ratio of specific emissivities per ion (see below). We found the mass ratios of $M_{\rm Mn}/M_{\rm Cr}$ = 0.10--0.46 and 0.18--0.66 in the entire remnant and the thermal dominant regions of the {\it Chandra} data, respectively. The $M_{\rm Mn}/M_{\rm Cr}$ in the {\it Suzaku} observation shows a much larger error of 0.08--0.84, being consistent with the Chandra measurements within the uncertainty.

\begin{figure*}[t!]
 \begin{center}
  \includegraphics[width=16cm, bb=0 0 970 451]{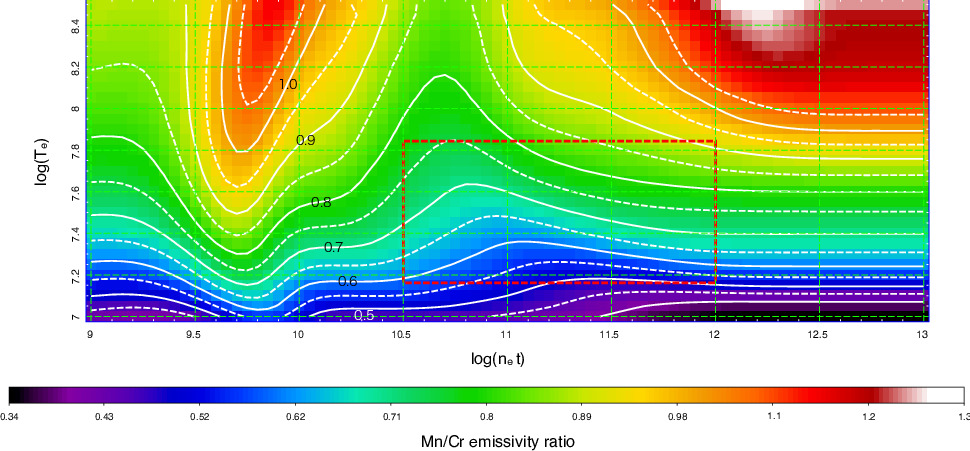}
 \end{center}
\caption{The Mn/Cr emissivity ratio map from the atomic data in \cite{2015ApJ...801L..31Y}. The broken red box shows the parameter range for Cassiopeia A; 1.25 keV $< kT_{\rm e} <$ 6 keV, $10^{10.5}$ cm$^{-3}$ s $< n_{\rm e}t <$ $10^{12}$ cm$^{-3}$ s. The color and contours show the emissivity ratios correspond to specific plasma parameters.}
\label{fig:f2}
\end{figure*}

For a range of the plasma parameters that are appropriate to Cassiopeia A, we conservatively constrain the range of the possible emissivity ratio of $\epsilon_{\rm Mn}/\epsilon_{\rm Cr}$ as 0.48--0.92. Figure \ref{fig:f2} shows the Mn/Cr emissivity ratio map and the plasma parameter range for Cassiopeia A. The X-ray spectra in Cassiopeia A have been well fitted by two temperature models \citep{2002A&A...381.1039W}. The hot plasma component is responsible for all the Fe-K emission and also dominates the continuum above 4 keV. Thus, the Cr-K ($\sim$5.63 keV) and Mn-K ($\sim$6.16 keV) lines may also come from the hot plasma component where their electron temperatures are in the range of 2 keV $< kT_{\rm e} <$ 6 keV.
On the other hand, the single-component models shown in \cite{2012ApJ...746..130H} indicated a lower electron temperature of 1.7 keV on average, and $kT_{\rm e} >$ 1.25 keV for most of the regions ($\sim$90\% in the area). Therefore, we assumed an enlarged temperature range of 1.25 keV $< kT_{\rm e} <$ 6 keV in this study. The range of ionization ages of $10^{10.5}$ cm$^{-3}$ s $< n_{\rm e}t <$ $10^{12}$ cm$^{-3}$ s is also consistent with all observations.

\begin{figure*}[t!]
 \begin{center}
  \includegraphics[width=16cm, bb=0 0 1088 511]{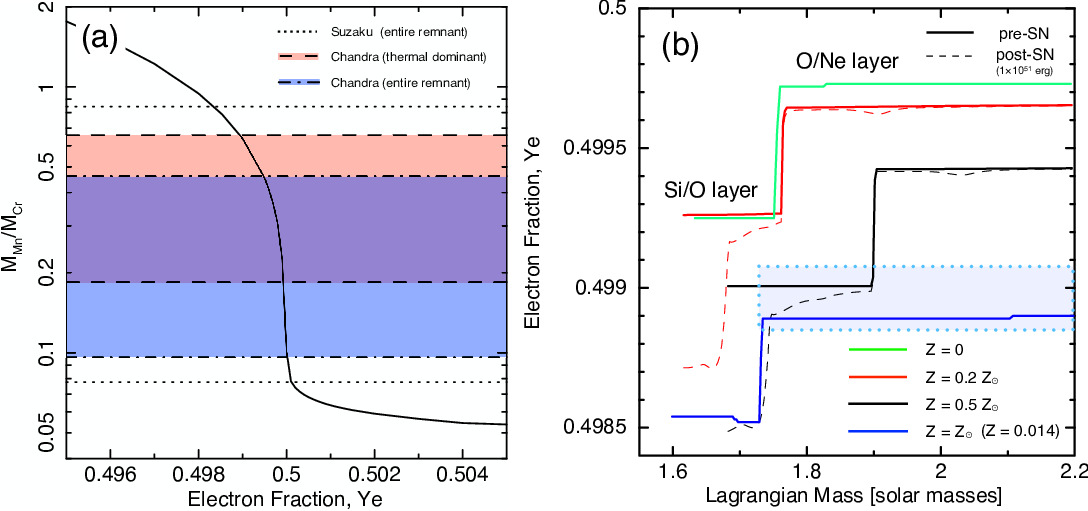}
 \end{center}
\caption{(a) $M_{\rm Mn}/M_{\rm Cr}$ vs. electron fraction at the incomplete Si burning layer. The faint red and blue areas show the $M_{\rm Mn}/M_{\rm Cr}$ ratios observed by {\it Chandra}. The range between the black dot lines indicates the $M_{\rm Mn}/M_{\rm Cr}$ ratios from {\it Suzaku}. 
The solid curve indicates the mass ratio of Mn/Cr obtained by the explosive nucleosynthesis calculations with a specific thermal profile as a function of electron fraction.
See the text for details of the thermal profile.
(b) The electron fraction in the several 15 $M_\odot$ progenitor stars. The green, red, black and blue solid lines show the $Y_
{\rm e}$ profiles of $Z = 0, 0.2 Z_\odot, 0.5 Z_\odot$ and $Z_\odot$ models, respectively. The broken lines show the $Y_{\rm e}$ profiles just after the explosion. The light blue area shows the range of $Y_{\rm e}$ at the O/Ne layer for different progenitor masses (13--20 $M_\odot$) of solar-metallicity stars \citep[models L in][]{2019ApJ...881...16Y}.}
\label{fig:f3}
\end{figure*}

We estimate that the electron fraction, $Y_{\rm e}$, at the incomplete Si burning layer in Cassiopeia A is $0.4990 \lesssim Y_{\rm e} \lesssim 0.5$ from the {\it Chandra} data.
Here, we calculated the explosive nucleosynthesis using a specific thermal profile with a peak temperature of 5.26 $\times 10^{9}$ K where the Cr yields are maximized \citep[similar to Fig. 4 in][]{2005ApJ...619..427U}.
This thermal profile comes from mass coordinate $M_r = 1.676 M_\odot$ in the supernova explosion model with an explosion energy of $E_{\rm exp} = 1 \times 10^{51}$ ergs and evolved from a solar-metallicity 15 $M_\odot$ star.
Details for the model are presented in the next section.
Figure \ref{fig:f3}(a) shows a relation between the $M_{\rm Mn}/M_{\rm Cr}$ ratio and the electron fraction for the incomplete Si burning regime. 
Interestingly, the inferred mass ratio is at $Y_{\rm e}$ $\sim$ 0.5, where the $M_{\rm Mn}/M_{\rm Cr}$ is very sensitive to the value of $Y_{\rm e}$. Therefore, we could tightly limit it even with the large errors in the derived mass ratio.

\section{Discussion on the Progenitor Metallicity and Explosion Energy}

There has been no observational constraint on the initial metallicity of the progenitor star of Cassiopeia A, even though it would be of great value in understanding the mass-loss history of the progenitor. We here estimate the progenitor's metallicity using our observational constraint on $Y_{\rm e}$ (from the Mn/Cr mass ratio) in Cassiopeia A for the first time. The value of $Y_{\rm e}$ at the incomplete Si burning layer estimated in the previous section constrains the degree of neutronization with the interior of the progenitor star. The main factor influencing the $Y_{\rm e}$ value in massive stars is the initial metallicity as well as the electron capture process during hydrostatic evolution (which causes the different $Y_{\rm e}$ values for the different compositional layers shown  in Fig.~\ref{fig:f3} (b)). We calculate several 1D core-collapse nucleosynthesis models from hydrodynamic evolution through explosion to compare with the observational results.

We calculate the evolution of 15 $M_\odot$ stars with metallicities of $Z = 0.014 (Z_\odot)$, 0.5$Z_\odot$, 0.2$Z_{\odot}$, and 0 from hydrogen burning until the central temperature reaches $10^{9.9}$ K.
For metallicities of $0.5Z_\odot$, we also calculate the evolution for 13, 17, and 20 $M_\odot$ stars.
These calculations are performed using a 1D stellar evolution code, HOngo Stellar Hydrodynamics Investigatar (HOSHI) code \citep{2016MNRAS.456.1320T,2018ApJ...857..111T,2019ApJ...871..153T,2019ApJ...881...16Y}.
Detailed parameter sets of the stellar models are the same as Set L in \citet{2019ApJ...881...16Y}.
Nucleosynthesis of 300 isotopic species is also calculated within the stellar evolution code.
The metallicity dependence of the mass loss rate is $Z^{0.85}$ for main-sequence stars and $Z^{0.5}$ for yellow and red supergiants \citep{2013A&A...558A.103G}.

Simulation of the supernova explosion is performed with a PPM hydrodynamics code \citep[e.g.,][]{1984JCoPh..54..174C,2005ApJ...619..427U}, assuming a spherically symmetric explosion.
The explosion energy $E_{\rm exp}$ is set in the range of (1, 2, 3, and 5)$\times 10^{51}$ ergs for $Z = 0.5Z_\odot$ and 0.2$Z_\odot$ stars and $1 \times 10^{51}$ergs for other metallicity stars.
The location of the mass cut is determined so that the ejected $^{56}$Ni mass is 0.07 $M_\odot$.
After the supernova explosion simulations, the explosive nucleosynthesis calculations are performed in a postprocessing step.
Radioactive decays in the supernova ejecta after 350 yr are also taken into account.

Figure \ref{fig:f3} (b) shows the $Y_{\rm e}$ profiles in the 15 $M_\odot$ progenitor models with different metallicities. In the He burning stage of stellar evolution, the decrease in $Y_{\rm e}$ is mainly attributed to the $\beta^+$ decay in the reaction sequence $^{14}$N($\alpha$,$\gamma$)$^{18}$F($\beta+$)$^{18}$O($\alpha$,$\gamma$)$^{22}$Ne \citep[e.g.,][]{1985ApJ...295..604T}. Thus, the $Y_{\rm e}$ depends on the initial CNO abundance (i.e., $^{14}$N abundance). This is similar to the process widely discussed in the context of SN Ia progenitors \citep[e.g.,][]{2003ApJ...590L..83T,2008ApJ...680L..33B,2013ApJ...767L..10P}. During carbon and oxygen 
burning, electron capture on nuclei is also effective. Thus, at the innermost shell (i.e., the Si/O layer in Figure \ref{fig:f3} (b)), the $Y_{\rm e}$ value shows a sharp decline.

In almost all previous studies, the progenitor has been assumed to have had solar metallicity \citep[e.g.,][]{2020NatAs.tmp....4K}. Interestingly, our tight upper limit of $Y_{\rm e} \gtrsim 0.4990$ at the incomplete Si burning layer is inconsistent with the low values of $Y_{\rm e}$ at all layers of a solar-metallicity progenitor (see the blue profile in Figure \ref{fig:f3} (b)). In addition, the Si/O layer has much smaller $Y_{\rm e}$ than that in the O/Ne layer. The large value of $Y_{\rm e}$ inferred in this study would thus suggest the O/Ne layer as the site for incomplete Si burning that produces the bulk of the Mn and Cr we see. However, for a nominal explosion energy of 1$\times$10$^{51}$ erg, the incomplete Si burning region is located at the Si/O layer in our calculations. For larger explosion energies, the Si-burning region can shift outward \citep[e.g.,][]{2001ApJ...555..880N}, which would help to reproduce the observational $Y_{\rm e}$. Previous studies have also provided evidence for larger than nominal explosion energies in the range  (1.5--4)$\times$10$^{51}$ erg for Cassiopeia A \citep[e.g.,][]{2003ApJ...593L..23C,2004NewAR..48...61V,2006ApJ...640..891Y,2016ApJ...822...22O,2017ApJ...842...13W}. Thus, a combination of low metallicity and large explosion energy would be appropriate for explaining the observational constraint on $Y_{\rm e}$ in Cassiopeia A. 

\begin{figure}[t!]
 \begin{center}
  \includegraphics[width=8cm, bb=0 0 771 728]{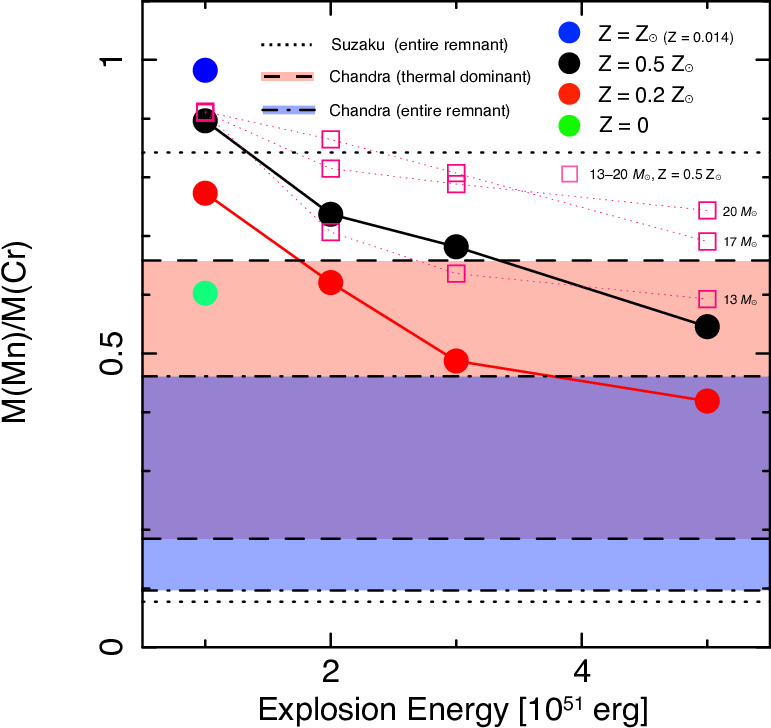}
 \end{center}
\caption{Relation between Mn to Cr mass ratio, $M_{\rm Mn}/M_{\rm Cr}$, and the explosion energy. The faint red and blue areas show the observed $M_{\rm Mn}/M_{\rm Cr}$ ratios by {\it Chandra}. The range between the black dotted lines indicates the $M_{\rm Mn}/M_{\rm Cr}$ ratios by {\it Suzaku}. The green, red, black and blue circles show the $M_{\rm Mn}/M_{\rm Cr}$ ratios (integrated over the entire ejecta) for the models with metallictites of $Z = 0, 0.2 Z_\odot, 0.5 Z_\odot$ and $Z_\odot$, respectively. For the filled circles, we assumed a 15 $M_\odot$ progenitor. The magenta open boxes show the $M_{\rm Mn}/M_{\rm Cr}$ ratios for different progenitor masses of 13, 17, and 20 $M_\odot$.}
\label{fig:f4}
\end{figure}

In Figure \ref{fig:f4}, we compare the observed $M_{\rm Mn}/M_{\rm Cr}$ ratio in Cassiopeia A with those in several one-dimensional nucleosynthesis models for the 15 $M_\odot$ progenitor. Here, we use the total ejected masses of Mn and Cr for estimating the mass ratio $M_{\rm Mn}/M_{\rm Cr}$ from the nucleosynthesis models. In fact, the $Y_{
\rm e}$ dependence on the Mn/Cr mass ratio estimated from the integrated ejecta is not exactly the same as in Figure \ref{fig:f3}(a). We find that a quite low metallicity of $Z < 0.2 Z_\odot$ at most is needed for reproducing the observed limits if we assume the normal explosion energy of 1$\times$10$^{51}$ erg. Increasing the explosion energy, the $M_{\rm Mn}/M_{\rm Cr}$ ratio in our models would fall within the observational range even with a higher metallicity of $0.5 Z_\odot$. The observational values for $M_{\rm Mn}/M_{\rm Cr}$ derived from the X-ray spectrum of the thermal dominated portion of the remnant compared to that from the entire remnant differ in whether or not the half-solar metallicity models are allowed for the highest explosion energy plotted.  Thus, to be conservative we accept the half-solar metallicity model as being consistent with the {\it Chandra} spectral data. In summary, we suggest that the progenitor metallicity is in the range of $0.2 Z_\odot \lesssim Z \lesssim 0.5 Z_\odot$ and the explosion energy should be at least 2$\times$10$^{51}$ erg, or even larger. 
In the case of Cassiopeia A's progenitor, a half-solar metallicity progenitor may be possible, however a progenitor with $Z = 0.2 Z_\odot$ would not be realistic \citep[see the green curve of the bottom panel of Figure 5 in][which corresponds to the metallicity distribution function: MDF around Cassiopeia A]{2015ApJ...808..132H}.

The $Y_{\rm e}$ value in the progenitor models also depends on the progenitor mass. The difference of the progenitor mass changes the position of the carbon-shell burning during  stellar evolution and changes the degree of neutronization in the O/Ne layer. Therefore, we also investigated the mass ratio $M_{\rm Mn}/M_{\rm Cr}$ dependence on the progenitor mass. The magenta boxes in Figure \ref{fig:f4} show the $M_{\rm Mn}/M_{\rm Cr}$ ratios in 13--20 $M_{\odot}$ models with half-solar metallicity. We found that only a few models fall into the allowed {\it Chandra} range (red area), although the $M_{\rm Mn}/M_{\rm Cr}$ dependence on the progenitor mass seems not to be negligible. We note that the $M_{\rm Mn}/M_{\rm Cr}$ constraint from the spectrum of the entire remnant does not formally allow the half-solar metallicity models for any of the range of progenitor masses shown with magenta boxes in Fig.~\ref{fig:f4}.  Again to be conservative we do not apply this constraint strictly.

We also investigate whether some solar-metallicity models with a different progenitor mass could produce the observed Mn/Cr ratio or not. We additionally calculated 13--20 $M_{\odot}$ progenitor models (without explosive nucleosynthesis) with solar metallicity to confirm it. We found the difference of $Y_{\rm e}$ in the O/Ne layer among them is not too significant (see light blue area in Figure \ref{fig:f3}$(b), \Delta Y_{\rm e} \sim 0.0002$). Here, the highest $Y_{\rm e}$ of $\sim0.4991$ in the O/Ne layers is close to that at the Si/O layer in the 15 $M_{\odot}$ progenitor with the half-solar metallicity, while the $Y_{\rm e}$ at the Si/O layer is still too low as $\sim0.4987$ at most. In this case, we need to put the entire Si-burning layer onto the O/Ne layer to obtain a low Mn/Cr ratio, which implies a quite high explosion energy ($\gtrsim 5 \times 10^{51}$ erg). Even if it were possible to put the entire Si-burning layer onto the O/Ne layer, the Mn/Cr ratio expected from its electron fraction would be at the upper limit of the {\it Chandra} measurements. Thus, we conclude that the current suite of spherically-symmetric, solar-metallicity models have difficulty reproducing the observation, even for different progenitor masses.


If there was a strong asphericity in the explosion, the shock may be sufficiently strong to have the incomplete Si at the O/Ne layer (suggesting an angle-dependent release of the explosion energy). Cassiopeia A is thought to have experienced an asymmetric explosion \citep[e.g.,][]{2000ApJ...528L.109H,2014Natur.506..339G}. On the other hand, it is difficult to discuss the asymmetric effect quantitatively using our one-dimensional models and need further multi-dimensional studies to confirm it. Therefore, we conservatively conclude that an energetic explosion ($>$2$\times$10$^{51}$ erg) and/or asymmetric explosion would be needed for Cassiopeia A. 

At present, the total $^{56}$Ni mass in Cassiopeia A has a large uncertainty \citep[0.058--0.4 $M_{\odot}$;][]{2006ApJ...640..891Y,2009ApJ...697...29E,2012ApJ...746..130H,2016ApJ...822...22O}. In our one-dimensional models, the total $^{56}$Ni mass of 0.07 $M_\odot$ is mainly determined by the location of the mass cut. On the other hand, the incomplete Si burning is not affected by the mass cut position, so we do not expect the Mn/Cr ratio to change much as the location of the mass cut is changed. To confirm this, we also calculated SN models that have a larger ejected $^{56}$Ni mass of 0.10 $M_{\odot}$ for $E_{\rm exp} = 5 \times 10^{51}$ ergs. In the case of the sub-solar metallicity model, the Mn/Cr ratio changes from 55\% ($^{56}$Ni mass of 0.07 $M_{\odot}$) to 57\% ($^{56}$Ni mass of 0.10 $M_{\odot}$), which is not a significant change. We conclude that the difference of the total $^{56}$Ni mass is not significant for our interpretation. 

The low metallicity for the progenitor would also have an impact on the yields of the odd-Z elements because the production of these elements occurs via reactions that rely on the presence of metals (e.g., $^{14}$N) in the burning layer that create a flux of neutrons. The odd-Z elements Na, Al, P, and K are produced in various nuclear burning phases during the lifetime of a massive star. Na and Al are synthesized in the hydrostatic carbon and neon-burning stages of massive stars, whereas P and K are synthesized in the explosive-burning stages. This implies that the yields of Na and Al are more sensitive to the progenitor mass than P and K. These elements are expected to be detected by future X-ray calorimeter missions, such as {\it XRISM} \citep{2014arXiv1412.1169H}. Thus, the information on the metallicity derived in the present work would hopefully lead to a comprehensive understanding of the Cassiopeia A progenitor, once it is coupled with future information from the analysis of the odd-Z elements in the remnant.

Recent multi-dimensional simulations indicate that the majority of the inner ejecta has $Y_{\rm e} \geq 0.5$ post explosion due to electron-neutrino trapping during the infall phase of the core collapse \citep[e.g.,][]{2020MNRAS.491.2715B}. This effect has been seen in other simulations \citep[see citations in][]{2020MNRAS.491.2715B} but remains provisional in lieu of better handling of neutrino transport in the simulations. If neutrino heating can extend to the Si/O layer or even the inner O/Ne layer (neither of which is clearly established yet), it would help to increase slightly the $Y_{\rm e}$ value at the incomplete Si burning layer over the values purely from hydrostatic evolution. Future insights on this topic from theory and multi-dimensional simulations would be valuable.



\section{A Type IIb Supernova with Sub-Solar metallicity?}

The light echo spectrum indicated that Cassiopeia A is a remnant of an SN IIb that had lost most of its hydrogen envelope prior to the explosion \citep{2008Sci...320.1195K}.
To strip its hydrogen envelope, mass loss processes such as binary interaction and stellar wind are important factors \citep[e.g.,][]{2010ApJ...725..940Y,2011A&A...528A.131C,2017ApJ...840...10Y,2019ApJ...885..130S}. Here, our new information on the initial metallicity in Cassiopeia A would be useful to understand the mass-loss history in the progenitor.

It has been generally accepted that the progenitor of Cassiopeia A most likely lost its hydrogen envelope via binary interactions \citep[e.g.,][]{2006ApJ...640..891Y,2008Sci...320.1195K}. On the other hand, \cite{2018MNRAS.473.1633K} and  \cite{2019A&A...623A..34K} argue that there is no surviving companion star in the remnant by setting tight upper limits on the brightness of possible candidates, raises a challenge to the binary-interaction scenario. In addition, the hypothesis of the single-star progenitor for Cassiopeia A can not be easily ruled out for now \citep{2019A&A...623A..34K}. 

Our new constraint on the initial metallicity of $Z \lesssim 0.5 Z_\odot$ could tightly limit the possibility of the single-star progenitor for Cassiopeia A.
In the single star evolution scenario, a high ZAMS mass of $M_{\rm ZAMS} \gtrsim 30 M_\odot$ would be needed even at solar metallicity \citep[e.g.,][]{2004MNRAS.353...87E,2011A&A...528A.131C} because the progenitor star needs to strip the hydrogen envelope away by its own stellar wind. At low metallicity, the mass-loss rate is lower, which means that a more massive progenitor is required. For example, in \cite{2019ApJ...885..130S}, only their most massive model of $M_{\rm ZAMS} \simeq 50 M_\odot$ with a sub-solar metallicity of $Z = 0.005$ ($= 0.36 Z_\odot$ in our case) could be barely labeled as a type IIb progenitor. However, such massive stars are not expected to successfully explode \citep{1999ApJ...522..413F}. In addition, such very massive stars ($\sim$50 $M_\odot$) do not seem to exist in the neighbourhood of Cassiopeia A \citep[see Appendix B in][]{2019A&A...623A..34K} and the estimated progenitor mass for Cassiopeia A \citep[= 15--25 $M_\odot$; ][]{2006ApJ...640..891Y,2016ApJ...822...22O} is much lower. The total ejecta mass has been estimated to be 2--4 $M_\odot$ \citep[e.g.,][]{2003A&A...398.1021W,2012ApJ...746..130H}, which also rules out a high mass progenitor above 18 $M_\odot$ \citep[e.g.,][]{2016ApJ...821...38S,2018ApJ...863..127K}. 
Although there are still large uncertainties in the wind-mass loss rate, it is very unlikely that a star with such a low mass and low metallicity ($M_{\rm ZAMS} \lesssim 20 M_\odot$ \& $Z \lesssim 0.5 Z_\odot$) will shed most of its hydrogen-rich envelope purely through stellar winds. Thus we conclude that a single-star progenitor is not suitable for the sub-solar metallicity progenitor of Cassiopeia A.

Binary stars, on the other hand, can produce SN IIb progenitors from much lower initial masses \citep[$\lesssim 20 M_\odot$; e.g.][]{2016MNRAS.457..328L,2017ApJ...840...90O,2017ApJ...840...10Y,2019ApJ...885..130S}, and lower metallicities are in fact favoured for this case \citep{2017ApJ...840...10Y,2019ApJ...885..130S}. The non-detection of a surviving companion narrows down the possible evolutionary channels.
The only possibilities left for Cassiopeia A's progenitor scenario are a binary with a compact-object companion (white dwarf, neutron star, or black hole), a binary merger \citep{1995PhR...256..173N}, or an ejected star from a disrupted binary system as discussed in \cite{2019A&A...623A..34K}. Among the compact object companion scenarios, only the black hole case can experience conservative mass transfer. This is because only the black hole companion can achieve a high enough mass ratio ($q \gtrsim$ 0.3) with the progenitor star of Cassiopeia A (15--20 $M_\odot$) to ensure conservative mass transfer. However, it is not trivial to understand how the black hole companion could have formed in the first place. In the case of a WD/NS companion, the mass ratio is small ($q\sim0.1$) and thus inevitably experiences non-conservative mass transfer. This will lead to a common envelope (CE) phase, where most of the hydrogen envelope will be ejected through a dynamical process and leaves a tight binary \citep[e.g.,][]{2013A&ARv..21...59I}. It is not clear whether any hydrogen can be left on the progenitor after CE phases. Even if there is any remaining hydrogen, it will most likely be removed before explosion via post-CE stellar winds or further mass transfer events because of the tight post-CE orbit. More theoretical investigations are required to further explore this channel. 

Stellar merger products have also been claimed to create stripped-envelope SNe \citep{1995PhR...256..173N,2017ApJ...842..125Z}. In those models, an unstable mass transfer phase leads to a stellar merger which unbinds part of the envelope. It is not clear what fraction of the envelope can be lost in the merging process, but hydrodynamical simulations suggest it is merely $<10\%$ \citep{1995ApJ...445L.117L,2019Natur.574..211S}. The rest of the envelope needs to be stripped through stellar winds, which is unlikely for low-metallicity stars.

In the disrupted-binary scenario discussed in \cite{2019A&A...623A..34K}, the system became unbound due to the first supernova in the system, before the Cassiopeia A explosion. The hydrogen envelope of the Cassiopeia A progenitor should have been lost through binary interactions before becoming unbound. According to some binary population synthesis studies, $\sim7.6\%$ of stripped-envelope supernovae arise in disrupted binaries \citep{2017ApJ...842..125Z}. Most of them are essentially single progenitors that have lost their envelope through stellar winds which would not apply to the case of Cassiopeia A based on our discussions above. There is a small contribution from systems which have experienced reverse mass transfer, where the secondary transfers mass back to the primary. This channel is strongly dependent on how much mass is retained during mass transfer and the assumptions made on angular momentum loss, but it is nevertheless a very rare case. 

Here, we raise another possibility where the Cassiopeia A progenitor lost its envelope and the binary became unbound at the same time. If the binary had an initial mass ratio close to unity $q\sim1$ and a wide enough separation, both stars can become red supergiants concurrently. Such stars have envelopes with very low binding energy, which can easily be blown off by supernova ejecta \citep[e.g.,][]{2014ApJ...792...66H}. If the separation was in the right range, the primary SN can both disrupt the binary \textit{and} remove a substantial amount of the companion's envelope. Then the secondary SN can appear as an SN IIb without a companion like Cassiopeia A. Further theoretical studies are needed to validate this scenario.


\begin{figure*}[t!]
 \begin{center}
  \includegraphics[width=17cm, bb=0 0 1032 554]{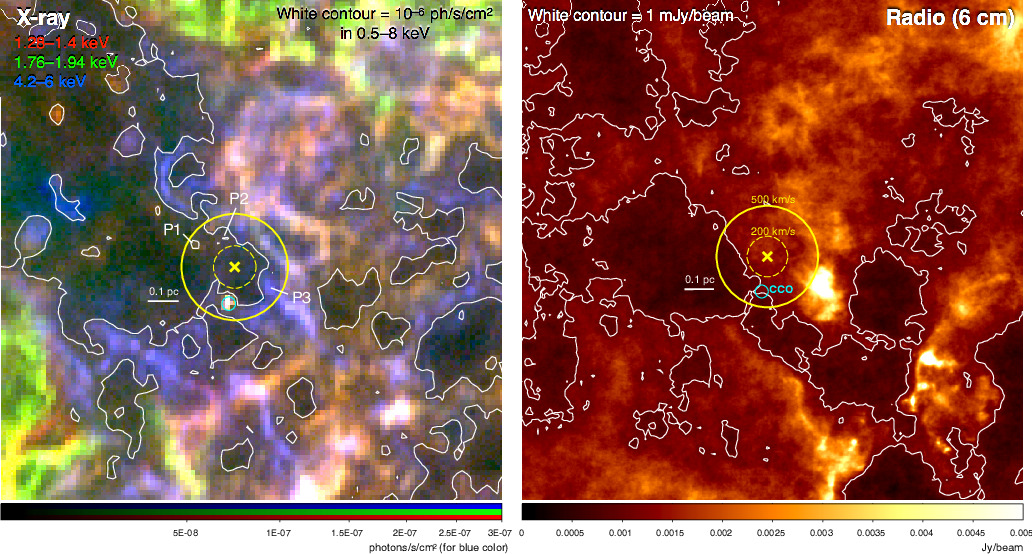}
 \end{center}
\caption{X-ray ({\it Chandra} in 2004 with $\sim1$ Msec: three color) and radio (VLA in 2000--2001: 6 cm) images around the explosion center of Cassiopeia A. In X-rays, red, green and blue show images in 1.28--1.4 keV (Mg K \& Fe L), 1.76--1.96 keV (Si K) and 4.2--6 keV (continuum), respectively. We use the explosion center derived from the optical observations \citep[$\alpha =$ 23$^{\rm h}$23$^{\rm m}$27$^{\rm s}_{\cdot}$77, $\delta = $58$^{\circ}$48$^{\prime}$49$^{\prime\prime}_{\cdot}$4;][]{2001AJ....122..297T,2006ApJ...645..283F}. The position of the central compact object (CCO) is marked by the cyan circle. Some knotty features around the center are labeled as P1, P2 and P3.}
\label{fig:f5}
\end{figure*}

\section{Isolated Black Hole in Cassiopeia A?}
If the black hole was the companion star of the progenitor star of Cassiopeia A as discussed in the previous section, an isolated-runaway black hole should exist around the explosion site. Such an isolated black hole may be able to be a radio/X-ray source \citep[e.g.,][]{2002MNRAS.334..553A,2005MNRAS.360L..30M,2018MNRAS.477..791T,2018MNRAS.475.1251M}. Thus, to search for the isolated black hole around the remnant will be important to verify the black-hole-companion scenario. 

We roughly estimate the luminosity $L$ of a black hole from the Bondi-Hoyle accretion \citep{1944MNRAS.104..273B} in a gas cloud around the explosion center of Cassiopeia A. At first, we estimate the Bondi-Hoyle accretion rate $\dot M$ as
\begin{eqnarray} \label{eq1}
\dot M &=&  \lambda \cdot 4 \pi \frac{(GM_{\rm BH})^2 \rho}{(v^2 + c_s^2)^{3/2}} \nonumber \\
&\approx& 1.9 \times 10^{9} ~{\rm g~s^{-1}}\nonumber \\
&\times& \left( \frac{\lambda}{0.01} \right) \left( \frac{M_{\rm BH}}{10~M_\odot} \right)^2 \left( \frac{\rho}{5~{\rm cm}^{-3}~m_p} \right) \left[ \frac{v^2 + c_s^2}{(100~{\rm km s}^{-1})^2} \right]^{-3/2}
\end{eqnarray}
where $G$, $M_{\rm BH}$, $\rho$, $v$, $c_s$ and $\lambda$ are the gravitational constant, the black hole mass, the gas mass density, the black-hole velocity relative to the ambient gas, the sound speed and the accretion efficiency, respectively. Here, the black hole must be moving in the unshocked ejecta of the remnant. Thus, the conditions of the unshocked ejecta (i.e., density and temperature) are critical parameters to consider the Bondi-Hoyle accretion. The low-frequency radio and near-infrared observations showed an electron density of 4.2 cm$^{-3}$ and a temperature of about 100 K in the unshocked ejecta \citep{2014ApJ...785....7D,2018ApJ...866..128R}. Therefore, the accretion rate is much smaller than that in molecular clouds due to the low density. The low temperature implies a low sound speed ($<$ 10 km s$^{-1}$). 
Assuming that the SN kick to the black hole is small, the velocity of the black hole is roughly determined by its orbital velocity prior to explosion ($v\sim\sqrt{GM/a}$), where $a$ is the separation of the binary system. The progenitor should have been filling its Roche lobe before the explosion so the pre-SN separation can be estimated as a few times the donor radius depending on the mass ratio \citep{1983ApJ...268..368E}. So if we assume the separation to be $a=300~R_\odot$ and a total system mass of $M=25~M_\odot$, the velocity can be estimated as $\sim130$~km~s$^{-1}$.

The accretion efficiency $\lambda$ of the isolated black holes is unclear. For example, in the case of active galactic nuclei (AGN), the efficiency varies from AGN to AGN, over the range of $\lambda =$ 0.1 to 0.001 where $\lambda =$ 0.01 is the typical value \citep{2005ApJ...624..155P}. Although it is not clear whether the accretion efficiency for the supermassive black holes can apply to the stellar mass black holes or not, we optimistically adapt the typical AGN's efficiency of $\lambda = 0.01$ in following calculations. 

Using the accretion rate, the bolometric luminosity is calculated as
\begin{eqnarray}
L &=& \epsilon \dot M c^2 = 1.7 \times 10^{29} \left( \frac{\epsilon}{0.1} \right) {\rm ~ erg ~ s}^{-1}
\end{eqnarray}
where $\epsilon$ is the radiation efficiency.  This is much fainter than the luminosity of the central compact object \citep[$L \sim$ 10$^{33-34}$ erg s$^{-1}$;][]{2000ApJ...531L..53P,2001ApJ...548..800C}. 
We here set the radiation efficiency as $\epsilon = 0.1$, however this assumption may be too optimistic. In the case of low accretion rates as in Eq. (\ref{eq1}), the radiation efficiency should be much smaller \citep[$\epsilon\sim10^{-10}$, e.g.,][]{1995ApJ...452..710N,2018MNRAS.477..791T}, which would make the detection of the isolated black hole impossible.
Nevertheless, in following discussions, we still continue to discuss the detectability with the optimistic radiation efficiency since we do not know actual radiation efficiency for isolated black holes.

Assuming the distance of $D =$ 3.4 kpc to the remnant \citep{2006ApJ...645..283F}, we obtain the flux as,
\begin{eqnarray} \label{eq3}
F &=& L/4\pi D^2 \nonumber \\
  &\approx& 1.3 \times 10^{-16} \left( \frac{\epsilon}{0.1} \right)  \left( \frac{D}{3.4~{\rm kpc}} \right)^{-2}  {\rm ~ erg ~ s^{-1} cm^{-2}} 
\end{eqnarray}
This is comparable to the {\it Chandra}'s sensitivity limit for X-ray point sources with a 1-Msec observation \citep{2005ApJS..161...21L}.

Figure \ref{fig:f5} shows X-ray and 6-cm radio\footnote{\url{http://homepages.spa.umn.edu/~tdelaney/cas/}} maps around the explosion center of Cassiopeia A. We found no obvious point source within the radius corresponding to the distance with a transverse velocity of 200 km s$^{-1}$ for 340 years (see broken circles in the figure). The X-ray surface brightness in 0.5--8 keV within the broken circle in Figure \ref{fig:f5} is $< 10^{-6}$ photons s$^{-1}$ cm$^{-2}$. The photon flux of $10^{-6}$ photons s$^{-1}$ cm$^{-2}$ corresponds to the unabsorbed energy flux of $\sim 1\times10^{-14}$ erg s$^{-1}$ cm$^{-2}$ in 0.5--8 keV, where we assumed an absorbed hard power law spectrum ($\Gamma = 1.5$, $N_{\rm H} = 1.5\times10^{22}$ cm$^{-2}$) to calculate it. In other words, the current data already has sensitivity to the isolated black hole whose flux is above $\sim 1\times10^{-14}$ erg s$^{-1}$ cm$^{-2}$, which is about 100 times higher than the estimated flux in Eq. (\ref{eq3}). At least, we can exclude the existence of a black hole whose mass is above 10 $M_\odot$ with efficient accretion and radiation ($\lambda \sim$ 1 and $\epsilon \sim$ 1) using the current data.

If we extend the search region to the radius that corresponds to the distance with a transverse velocity of 500 km s$^{-1}$ for 340 years (see solid circles in Figure \ref{fig:f5}), we find some knotty structures within the circle. For example, we labeled three knotty structures as P1, P2 and P3 in Figure \ref{fig:f5}. The P1 and P2 are adjacent to each other, and their reddish colors are similar to each other. On the other hand, the P3 has a blue color that implies a hard X-ray spectrum of this source. We found that the P1 and P3 spectra could be well fitted with an absorbed non-equilibrium ionization (NEI) plasma model ($kT_{\rm e} \sim 0.9$ keV) and an absorbed power law model ($\Gamma \sim $ 2.4), respectively. It is most likely that the P1 and P2 knots come from the shocked ejecta and knot P3 is part of the non-thermal filaments \cite[e.g.,][]{2018ApJ...853...46S}. Thus, we conclude that these structures are not related to the isolated black hole. 


Even with {\it Chandra}'s long observation, it is impossible reach a definitive conclusion on the existence of such a compact object in Cassiopeia A. Further investigations by proposed future missions with significantly higher sensitivity for X-ray point sources \citep[e.g., {\it AXIS}, {\it FORCE};][]{2019BAAS...51g.107M,2016SPIE.9905E..1OM,2018SPIE10699E..2DN} may be helpful in revealing the existence of an isolated compact companion in this remnant.

\section{Summary}
Cassiopeia A is a special galactic object with unique potential to aid in understanding the explosion mechanism of stripped-envelope core-collapse supernovae. In particular, the bright X-ray flux of the remnant that makes it possible to detect emission from rare metal species also provides us detailed information on the explosive nucleosynthesis process. In this study, we discuss the degree of neutronization at the incomplete Si burning regime based on the first detection of manganese in the remnant.

We determined that the archival {\it Chandra} data of Cassiopeia A indicate a low Mn/Cr mass ratio of 0.10--0.66, which requires an electron fraction of 0.4990 $\lesssim Y_{\rm e} \lesssim$ 0.5  at the incomplete Si burning layer. We also found a consistent value using the {\it Suzaku} X-ray spectrum, which shows the robustness of our measurements. Comparing with several nucleosynthesis models for CC SNe, we conclude that an energetic explosion ($>$ 2$\times$10$^{51}$ erg) and/or asymmetric explosion of a sub-solar metallicity progenitor ($Z \lesssim 0.5 Z_\odot$) would be needed to reproduce such a high $Y_{\rm e}$ value.

The low metallicity of $Z \lesssim 0.5 Z_\odot$ along with previous observational constraints disfavors the single-star progenitor scenario and strongly support the binary-stripped progenitor for Cassiopeia A. On the other hand, the absence of the companion star still leaves problems for the binary-stripped scenario. We will need further studies for testing the surviving scenarios (a compact-object companion, a binary merger or a disrupted binary) to understand the mass-loss history in the progenitor system for Cassiopeia A.

\acknowledgments
We thank Adam Burrows and Shinya Yamada for helpful comments and discussions.
T.S.\ was supported by the Japan Society for the Promotion of Science (JSPS) KAKENHI grant Nos.\ JP19K14739, the Special Postdoctoral Researchers Program and FY 2019 Incentive Research Projects in RIKEN. K.M.\ was supported in part by the Grants-in-Aid for the Scientific Research of Japan Society for the Promotion of Science (JSPS, Nos. JP17H01130). S.N. is partially supported by the Grants-in-Aid for "Scientific Research of JSPS (KAKENHI) (A) 19H00693", Pioneering Program of RIKEN for Evolution of Matter in the Universe (r-EMU), and Interdisciplinary Theoretical and Mathematical Sciences Program of RIKEN (iTHEMS). J.P.H.\ acknowledges support for X-ray studies of SNRs from NASA grant NNX15AK71G to Rutgers University. Y.M.\ discussed the Suzaku data with Prof.\ H.~Tsunemi and other colleagues and expresses his thanks to them. We also thank the anonymous referee for comments that helped us to improve the manuscript.


\bibliography{sample63}{}
\bibliographystyle{aasjournal}



\end{document}